\documentclass[aps,prl,reprint,showpacs,amsmath,amssymb,superscriptaddress,floatfix]{revtex4-1}
\usepackage{bm}
\usepackage{bbold}
\usepackage{graphicx}
\usepackage{braket}

\begin{document}

\title{Sliding Luttinger liquid with alternating interwire couplings}
\author{S Begum}
\affiliation{Aston University, School of Engineering \& Applied Science -  Birmingham B4 7ET, UK}
\author{V Fleurov}
\affiliation{Raymond and Beverly Sackler Faculty of Exact Sciences, School of Physics and Astronomy, Tel-Aviv University -  Tel-Aviv 69978, Israel}
\author{V Kagalovsky}
\affiliation{Shamoon College of Engineering - Beer-Sheva 84105, Israel}
\affiliation{Max-Planck-Institut f\"ur Physik komplexer Systeme, N\"othnitzer Str., Dresden, Germany}
\author{I V Yurkevich}
\affiliation{Aston University, School of Engineering \& Applied Science -  Birmingham B4 7ET, UK}

\vspace{10pt}

\begin{abstract}
We study a phase diagram for the sliding Luttinger liquid (SLL) of coupled one-dimensional quantum wires packed in a two-dimensional array in the absence of a magnetic field. We analyse whether nearest-neighbour inter-wire interactions, stabilise the SLL phase. We construct an analogue of a Su-Schriefer-Heeger (SSH) model (allowing alternating couplings between wires). Calculating the scaling dimensions of the two most relevant perturbations, charge-density wave, and superconducting inter-wire couplings, but excluding the inter-wire single-particle hybridisation, we find a finite stability region for the SLL. It emerges due to the inter-wire forward scattering interaction, and remains stable up to a significant asymmetry between alternating couplings.
\end{abstract}

\vspace{2pc}

\maketitle

\section{Introduction}

The behavior of interacting electrons in 1d systems (Tomonaga- Luttinger model \cite{T50,L63}) has many features that differ them from electrons in 2d and 3d systems. The latter form Fermi liquid (FL) described by the Landau theory, which instead of strongly interacting fermions considers low energy quasiparticles retaining their Fermi statistics. This approach does not work in one dimensional systems, where geometrical constraints impose strong limitations on the electron-electron scattering. As a result the bosonic multielectron excitations play the dominant role \cite{H81}. A possibility of observing such a non-Fermi liquid behavior in higher dimensional systems was discussed in Ref. \cite{A90} with the aim to explain the unusual physics of high $T_c$ superconductors. Later on the strategy for creating a non Fermi liquid state, put forward in Refs. \cite{HLT99,EFKL00,MKL01,KML02,VC01,MK01}, was to consider a highly anisotropic 2D or 3D array of parallel identical periodically arranged quantum wires, coupled by forward scattering interactions. The electron liquid in these setups may form a so called Sliding Luttinger liquid phase (SLL), which was supposed to exhibit low energy physics similar to that of individual wires. It was expected that the conductivity of multi-channel (quasi-1D) strongly correlated systems demonstrated power law temperature dependence, which together with some other unusual properties was really observed in a number of strongly anisotropic systems \cite{BCLRSBM99,ZPBD99,BJGMBS01,KTAO04,KSCTVNA09,ZZ00,VHK06,C11,A04}.  These measurements were carried out at relatively high temperatures, where single particle and many particle processes, which potentially are capable of destroying the SLL state, were rather weak. The single particle back scattering in the individual wires \cite{G04,W90} can block the flow at low temperatures. That is why these wires must be clean of any defects which may cause scattering. There are also three interwire mechanisms which, when RG (renormalization group) relevant, can also destroy the SLL phase. These mechanisms are: single electron tunneling, which converts the SLL state into a FL state; two types of the multielectron interwire process can result either in a superconducting (SC)\cite{NYWLHYDZYTZ14} or charge density wave (CDW)\cite{SHZZ04} states.

Recently the problem of quantum wires coupled by Coulomb interaction under the condition that single electron interwire tunnelings are suppressed, was addressed in Ref. \cite{SY17}. A study was carried out of SLL in a stack of wires arranged in a lattice under the assumption that the Coulomb repulsion between electrons of the neighboring wires is dominant. This Coulombic SLL appears to be unstable towards formation of a CDW. It was also emphasized in \cite{FKLY18} that all three mechanisms always lead to the instability and it is impossible to create realistic conditions when all of them are irrelevant. It was also outlined that SLL state can be realized in the systems where both single electron and SC tunneling are effectively suppressed.

In this paper we consider a 2D system of parallel quantum wires separated by barriers of alternating transparencies in analogy to the Su-Schriefer-Heeger (SSH) model \cite{HKSS88}. We calculate analytically the scaling dimensions of CDW and SC inter-wire perturbations and find that there exists a finite region of stability of SLL, where both perturbations are irrelevant, and Luttinger liquid description is valid for a two-dimensional system.

\section{Model}

The system of $N$ parallel quantum wires is described by the standard bosonized Lagrangian
\begin{equation}\label{Lagrangian}
  {\cal L} = \frac{1}{4\pi} \left[\partial_t\varphi \partial_x\theta + \frac{1}{2} \partial_x \mbox{\boldmath$\theta$}^T \textsf{V}_\theta \partial_x \mbox{\boldmath$\theta$} + \frac{1}{2}\partial_x \mbox{\boldmath$\varphi$}^T  \textsf{V}_\varphi   \partial_x \mbox{\boldmath$\varphi$}\right].
\end{equation}
written in terms of the fields
$$
\begin{array}{ccc}
  \mbox{\boldmath$\theta$} & = & \{\theta_1,\theta_2, \cdots \theta_N\} \\
  \mbox{\boldmath$\varphi$} & = & \{\varphi_1, \varphi_2, \cdots \varphi_N\}.
  \label{fields}
\end{array}
$$
These fields represent the density $\delta\rho_i= \frac{1}{\pi} \partial_x \varphi_i$ and current $j_i = \frac{1}{\pi} \partial_x \theta_i$ fluctuations. The matrices $\textsf{V}_\varphi$ and $\textsf{V}_\theta$ are tri- diagonal with the principal diagonal elements expressed in terms of velocity $v$ and Luttinger parameter $K$, independent the number $i$ of the wire, i.e. all the wires are identical. These parameters are in their turn expressed in terms of forward $g_4$ and back $g_2$ scattering amplitudes. These two parameters are assumed to be equal. We also use the units in which the Fermi velocity $v_F =1$. As for the off-diagonal terms they alternate so that matrix elements connecting the wire $i$ with $i + 1$ for odd $i$ are also equal each other but differ from those with even $i$. As a result the density-density interaction matrix takes the form,
\begin{equation}\label{matrix}
\widehat{V}_{\phi} =
$$$$
\left(
  \begin{array}{ccccccc}
    1+  g &  g_o & 0 & 0 & \cdots & 0 & 0 \\
    g_o & 1 +  g & g_e & 0 & \cdots & 0 & 0 \\
    0 & g_e & 1 + g & g_e & \cdots & 0 & 0 \\
    0 & 0 & g_o & 1 + g & \cdots & 0 & 0 \\
    \cdots & \cdots & \cdots & \cdots & \cdots & \cdots & \cdots \\
    0 & 0 & 0 & 0 & \cdots & g_o & 1 + g  \\
  \end{array}
\right)
\end{equation}
with $g= 2\pi(g_4 + g_2)$. Below, we will analyze in detail the commonly accepted model that includes only a density-density neglecting the current -- current inter-wire interaction. Hence the current-current interaction matrix $\widehat{V}_{\theta}=\hat{\mathbb{1}}$ in Eq.~\eqref{Lagrangian}.

It is convenient to define a Luttinger matrix $\widehat{K}$ \cite{Yur1, Yur2, KLY, Yur, ACh, JLY, ACh2}  which is a generalization of the Luttinger parameter $K$ of a single channel. All scaling dimensions of all symmetry allowed perturbations can be expressed using this single matrix ${\widehat K}$. This matrix provides information on the relevance of the perturbations and, therefore, on the stability region of the SLL phase. In our case the $\widehat K$-matrix is the solution of the matrix equation $\widehat{K} \widehat{V}_{\phi} \widehat{K}=\hat{\mathbb{1}}$, i.e. the square root of the interaction matrix $ \widehat{K} = \widehat{V}_{\phi}^{-1/2}$.

\section{Eigenfunctions of interaction matrix}

In order to find $\widehat{K}$ we need to know the eigenvalues and eigenvectors of the matrix $\widehat{V}_\phi$. The equations for the eigenvectors $\psi$ and eigenvalues $\lambda$ of the matrix $\widehat{V}_{\phi}$ read
\begin{eqnarray}
g_o\,\psi(2j-1) + g_e\,\psi(2j+1)=\epsilon\,\psi(2j)\,,\nonumber \\
\quad j=1,...,M\,;\\
g_e\,\psi(2j-2) + g_o\,\psi(2j)=\epsilon\,\psi(2j-1)\,,\nonumber \\
\quad j=1,...,M+1\,,
\end{eqnarray}	
where we define $\epsilon = \lambda-(1+g)$. The second equation is extended by including the sites $0$ and $2M+2$, which satisfy the boundary conditions	
\begin{eqnarray}\label{bc}
\psi(0)=\psi(2M+2)=0\,,
\end{eqnarray}
We will seek eigenfunctions labelled by the wave vector $k$,
\begin{eqnarray}
\psi_k(2j)=e_k\,e^{ikj}+e_{-k}\,e^{-ikj}\,;\label{evenfun}\\
\psi_k(2j-1)=o_k\,e^{ikj}+o_{-k}\,e^{-ikj}\,.\label{oddfun}
\end{eqnarray}
The coefficients $e_k$ and $o_k$ satisfy the equations
\begin{eqnarray}
{\bar g}_k\,e_k&=&\epsilon_k\,o_k\,,\quad\quad \,\label{evenfun-1}\\
g_k\,o_k&=&\epsilon_k\,e_k\,,\label{oddfun-1}
\end{eqnarray}
where $g_k=g_o + g_e\,e^{ik}$ and overline denotes the complex conjugation.

The boundary conditions Eq.~(\ref{bc}) presented in the form
\begin{eqnarray}
e_{-k}=-e_k\,,\quad {\rm and} \quad \sin k(M+1)=0\,,\quad
\end{eqnarray}
impose quantisation of the wave vector
\begin{eqnarray}
k=\frac{\pi l}{M+1}\,,\quad l=1,...,M\,.
\end{eqnarray}

It follows from Eqs. (\ref{evenfun-1}) and (\ref{oddfun-1}) that the eigenvalues form two bands
\begin{eqnarray}
\epsilon_k^{\pm}=\pm\,|g_k|\,.
\end{eqnarray}
whereas the corresponding $2M $ normalized eigenvectors are:
\begin{eqnarray}
\psi_k^{\pm}(2j)&=&\frac{1}{\sqrt{M+1}}\,\sin kj\,,\nonumber\\
j = 1, \dots , M\\
\psi_k^{\pm}(2j-1)&=&\pm\frac{1}{\sqrt{M+1}}\,\sin (kj-\phi_k)\,,\\
j = 1, \dots ,\ M + 1,  \nonumber
\end{eqnarray}
where the half phase of reflection coefficient of the 'odd' component is defined by the equation
\begin{eqnarray}\label{def}
g_k=g_o + g_e\,e^{ik}=|g_k|\,e^{i\phi_k}\,.
\end{eqnarray}

Apart from the above $2 M $ bulk eigenvectors labelled by the wave vector $k$, there is one more solution exactly at $\epsilon = 0$. It is exponentially localised at the left boundary for $g_o < g_e$
\begin{eqnarray}
\psi_0(2j-1)=N_0^{-1/2} \left(-\frac{g_o}{g_e}\right)^j\,,\quad \psi_0(2j)=0\,,
\end{eqnarray}
where
$$
N_0=\frac{1-(-g_o/g_e)^{M+1}}{1 + g_o/g_e}
$$
and a similar solution, exponentially localized near the right boundary, exists at $g_o > g_e$.

As we have mentioned above $\widehat{K}= \widehat{V}_{\phi}^{-1/2}$. The diagonalization of the interaction matrix, carried out above, allows us to obtain the Luttinger matrix in the form
\begin{eqnarray}\label{K-matrix}
\widehat{K}_{ij} = K \left[\sum_{k,\sigma=\pm} \Lambda_{k,\sigma}^{-1/2}\,\psi^{\sigma}_k(i)\psi^{\sigma}_k(j) + \psi_0(i)\psi_0(j)\right]\,. \label{CDW}
\end{eqnarray}
where we made use of the eigen values
\begin{eqnarray}
\lambda_{k,\pm}= \frac{1}{K^2}\Lambda_{k,\pm} = \pm |g_k| + 1 + g\\
\quad{\rm and}\quad \lambda_0 = 1 + g \equiv K^{-2}\,, \nonumber
\end{eqnarray}
and the corresponding eigenvectors $\psi_k^{\pm}(i)$ and $\psi_0(i)$. We can readily write equation for the inverse Luttinger matrix
\begin{eqnarray}\label{K-1-matrix}
(\widehat{K}^{-1})_{ij} = K\left[\sum_{k,\sigma=\pm} \Lambda_{k,\sigma}^{1/2}\,\psi^{\sigma}_k(i)\psi^{\sigma}_k(j) + \psi_0(i)\psi_0(j)\right]\,.\label{SC}
\end{eqnarray}

We are looking for the situation when the SLL phase is stable with respect to two processes: formation of CDW and SC states. For this we have to make sure that the corresponding processes are RG irrelevant.   The scaling dimensions of couplings between the wires $i$ and $j$ are
\begin{equation}\label{cdwdim-1}
\Delta^{CDW}_{ij} = \widehat{K}_{ii} + \widehat{K}_{jj} - 2 \widehat{K}_{ij}
\end{equation}
and
\begin{equation}\label{csdim-1}
\Delta^{SC}_{ij} = (\widehat K^{-1})_{ii} + (\widehat K^{-1})_{jj} - 2 (\widehat K^{-1})_{ij}
\end{equation}
for the processes responsible for the CDW and SC phases formation respectively. We consider here only the most dangerous among them, corresponding to neighboring wires, so that $j = i + 1$.

\subsection{Bulk scaling dimensions}

Now we have to substitute the known eigenvalues and eigenfunctions into Eqs. (\ref{K-matrix}) and (\ref{K-1-matrix}) and then into Eqs. (\ref{cdwdim-1}) and (\ref{csdim-1}). After ordering the terms we get two CDW scaling dimensions
\begin{eqnarray}
\quad\,\Delta_e^{CDW} = \nonumber\\
2K \int_0^\pi \frac{dk}{\pi} \left[ \lambda_+^{-1/2} \sin^2\frac{\phi}{2} + \lambda_-^{-1/2} \cos^2 \frac{\phi}{2}\right]\label{De1}\\
\Delta_o^{CDW} = \nonumber\\
2K \int_0^\pi \frac{dk}{\pi} \left[ \lambda_+^{-1/2} \sin^2\frac{\phi - k}{2} + \lambda_-^{-1/2} \cos^2 \frac{\phi -k}{2}\right]\label{De2}
\end{eqnarray}
and two SC scaling dimensions
\begin{eqnarray}
\Delta_e^{SC} =  \nonumber \\
2K \int_0^\pi \frac{dk}{\pi} \left[ \lambda_+^{1/2} \sin^2\frac{\phi}{2} + \lambda_-^{1/2} \cos^2 \frac{\phi}{2}\right], \label{Do1}\\
\Delta_o^{SC} =  \nonumber \\
2K \int_0^\pi \frac{dk}{\pi} \left[ \lambda_+^{1/2} \sin^2\frac{\phi - k}{2} + \lambda_-^{1/2} \cos^2 \frac{\phi -k}{2}\right] \label{Do2}
\end{eqnarray}
These equations for scaling dimensions are valid far from the edges in the limit $M\to \infty$, since we have neglected exponentially decaying edge terms and averaged strongly oscillating terms at large $i$. The index $i$ may correspond to the even or odd wire with alternating coupling constants. That is why we get two CDW scaling dimensions and SC dimensions.

It is convenient to introduce two inter-channel parameters (instead of $g_{e,o}\tilde = g\mp\Delta $), i.e the relative strength of the inter-channel interaction $\alpha=\frac{2\tilde g}{1+g}$ and the relative strength of asymmetry $\beta = \frac{\Delta}{\tilde g}$, i.e. modulation. The two parameters $K$, $\alpha$, vary between $0$ and $1$. In principle, $\beta$ varies between -1 and 1, but as we'll see below the symmetry allows us to reduce the range for $\beta$ also to $(0,1)$. Now we can use the relations
\begin{eqnarray}
\lambda _{k}^{\pm}=1\pm\alpha\sqrt{\cos^2 \frac{k}{2}+\beta^2\sin^2 \frac{k}{2}}\\\nonumber
\cos\phi_k =\frac{\cos^2 \frac{k}{2}-\beta\sin^2 \frac{k}{2}}{\sqrt{\cos^2 \frac{k}{2}+\beta^2 \sin^2 \frac{k}{2}}}\\ \nonumber
\cos (\phi_k -k) =\frac{\cos^2 \frac{k}{2}+\beta\sin^2 \frac{k}{2}}{\sqrt{\cos^2 \frac{k}{2}+\beta^2 \sin^2 \frac{k}{2}}}\\
\sin\phi_k = \frac{\cos \frac{k}{2}\sin\frac{k}{2}(1 + \beta)}{\sqrt{\cos^2 \frac{k}{2}+\beta^2 \sin^2 \frac{k}{2}}}
\end{eqnarray}
In each pair of scaling dimensions one of the dimensions is more dangerous and it is used to get integrals (\ref{CDW-1}) and (\ref{SC-1}). We notice that due to the two following inequalities $\lambda _{k}^{+}>\lambda _{k}^{-}$ , and $\cos (\phi_k -k)>\cos (\phi_k )$, we have to use integrals in Eqs.(\ref{De1}) and (\ref{Do2}) as more dangerous (smaller), respectively (for $\beta >0$). For $\beta < 0$ the integrals in each pair interchange. As a result we get equations for the scaling dimensions, which are even in $\beta$. Therefore it is sufficient to consider $\beta$ varying from 0 to 1. Finally, we represent scaling dimensions in the form
\begin{equation}\label{scaling}
\Delta_{CDW} = 2\frac{K}{K_{CDW}},\ \ \ \Delta_{SC} = 2 \frac{K_{Sc}}{K}
\end{equation}
where
\begin{widetext}
\begin{eqnarray}
  K_{CDW}^{-1} &= & \int_0^{\pi/2} \frac{dk}{\pi} \left[\frac{1}{\sqrt{1 + \alpha r}} \left(1 - \frac{\cos^2k - |\beta| \sin^2 k}{r}\right) + \frac{1}{\sqrt{1 - \alpha r}} \left(1 + \frac{\cos^2k - |\beta| \sin^2 k}{r}\right) \right]\label{CDW-1}\\
K_{SC} &= & \int_0^{\pi/2} \frac{dk}{\pi} \left[\sqrt{1 + \alpha r} \left(1 - \frac{\cos^2k + |\beta| \sin^2 k}{r}\right) + \sqrt{1 - \alpha r} \left(1 + \frac{\cos^2k + |\beta| \sin^2 k}{r}\right) \right] \label{SC-1}
\end{eqnarray}
\end{widetext}
where $r = \sqrt{\cos^2 k + \beta^2 \sin^2 k }$.

\begin{figure}
\centering
\includegraphics*[width=1.2\columnwidth]{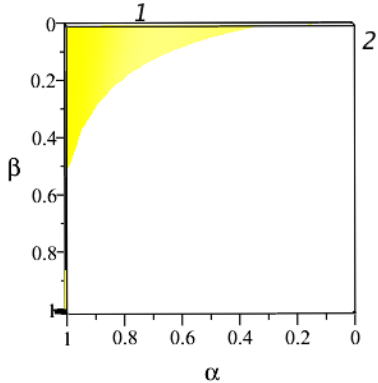}
\vspace{-5cm}
\caption{(Color online) A 3D plot of the product $\Pi$, Eq. (\ref{inequality}), showing the surface 1 (yellow online) crossed by the flat surface 2 (white online) - view from above. The region of stability of the SLL phase is clearly seen.}
\label{risunokVF-a}
\end{figure}

The interactions responsible for transitions to CDW or SC phases are both RG irrelevant if both scaling dimensions (\ref{scaling}) are larger than 2. It means that the intrawire Luttinger parameter $K$ must satisfy the condition
\begin{equation}\label{conditions}
K_{SC} > K > K_{CDW}.
\end{equation}
It follows from equation (\ref{conditions}) that it can be always satisfied (by choosing a proper Luttinger parameter $K$) if the product of integrals defined in equations (\ref{CDW-1}) and (\ref{SC-1}) is larger than 1:
\begin{equation}\label{inequality}
\Pi \equiv K_{SC} K_{CDW}^{-1} > 1.
\end{equation}
The inequality (\ref{inequality}) allows us to find straightforwardly the region of stability of the SLL phase. Figure \ref{risunokVF} shows a 3D graph of the product of integrals $\Pi$ (\ref{inequality}) as a function of the two parameters $\alpha$ and $\beta$ crossed by the flat surface drawn at level 1. There is a sector in the $\alpha - \beta$ plane where the product $\Pi$ is above the flat plane, i.e., $\Pi > 1$. This is the area where $K$ values, satisfying inequality (\ref{inequality}), can be found. At these values of $(\alpha,\beta)$ one can find the value of the third parameter $K$ such that both CDW and SC processes are RG irrelevant and the SLL phase is stable. This area is clearly exposed in the view from above of the same 3D plot in Fig. \ref{risunokVF-a} .

\begin{figure}\label{risunokVF}
\centering
\includegraphics*[width=1.2\columnwidth]{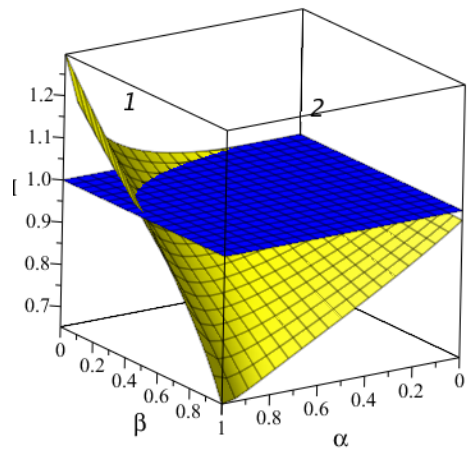}
\vspace{-5cm}
\caption{(Color online) A 3D plot of the product $\Pi$, Eq. (\ref{inequality}), showing the surface 1 (red online), which is crossed by the flat surface 2 (blue online). The part of the surface 1 residing above the surface 2 corresponds to a sector in $(\alpha,\beta)$ plane where the SLL phase is stable. }
\label{risunokVF}
\end{figure}
\begin{figure}
\centering
\includegraphics*[width=1.2\columnwidth]{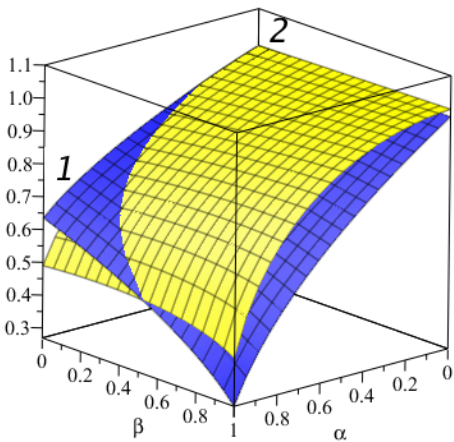}
\vspace{-5cm}
\caption{(Color online) $\alpha$ - $\beta$ - $K$ diagram. Surface 1 (blue online) corresponds to $K^{-1}_{CDW}(\alpha, \beta)$, and surface 2 (yellow online) corresponds to $K_{SC}(\alpha,\beta)$. There is a sector in $(\alpha, \beta)$ (the same as in Figure \ref{risunokVF}) where the surface 1 is above the surface 2 and stable SLL solutions are possible. The corresponding $K$ values lie between these two surfaces.}
\label{risunokVF1}
\end{figure}

The stability area has a shape close to a right triangle with catheti of about 0.5 along the $\beta$ axis and 0.5 along the $ \alpha $ axis. It occupies an ample share of the total $\alpha-\beta$ phase space - close to 10$\%$. In order to estimate at what values of the Luttinger parameter $K$ the SLL phase remains stable with respect to the formation of CDW  and SC phase we prepare another 3D plot, which shows the surface 1 (blue online) for the function $K^{-1}_{CDW}(\alpha,\beta)$,  and surface 2 (yellow online) for the function $K_{SC}(\alpha, \beta)$. The inequality  (\ref{conditions}) can be fulfilled if the surface 1 lies above the surface 2, and the desirable $K$ values must lie between these surfaces. One can readily see from Fig. \ref{risunokVF1} that the maximal span from 0.5 to 0.65 of the allowed $K$ values is at $\alpha =1,\ \beta =0$. Then the span gradually narrows and goes to zero with the increasing $\beta$ or decreasing $\alpha$.

Fig. \ref{risunokVF1} allows us also to address the opposite question - what happens when both CDW and SC perturbations are relevant. This happens when inequality signs in Eq. (\ref{conditions}) are inverse. In that case the perturbations compete, and in addition to the scaling dimensions, the initial values of perturbation amplitudes play an important role.

\begin{figure}
\centering
\includegraphics*[width=1.2\columnwidth]{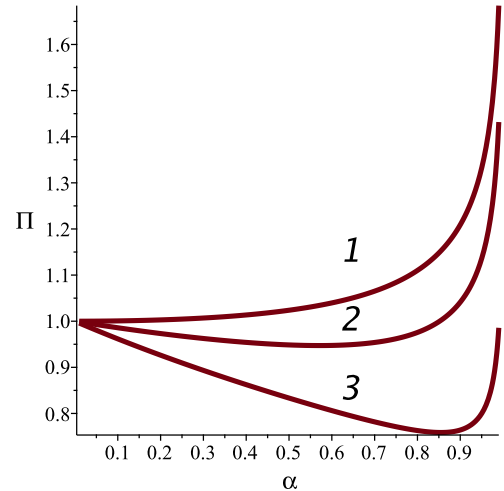}
\vspace{-5cm}
\caption{The plot of the product $\Pi$ as function of the parameter $\alpha$ for three values of the modulation parameter $\beta$; (1) - $\beta =0$; (2) - $\beta = 0.3$; (3) - $\beta = 0.8$.}
\label{risunokVF2}
\end{figure}

One can readily see that integrals (\ref{CDW-1}) and (\ref{SC-1}) equal one at $\alpha = 0$, i.e. $\Pi = 1$.  It means that the span of allowed values of $K$ reduces to zero and SLL phase is marginally stable only in the point at $K=1$. There is a stability stripe going from the stability triangle along the $\alpha$ axis up to this point. The stripe is so narrow that it cannot be seen in Figs. \ref{risunokVF} or \ref{risunokVF-a}. On the other hand the integral (\ref{CDW-1}) diverges at its low limit at $\alpha \to 1$. It means that there is also a very narrow stripe of stability going parallel to the $\beta$ axis along the line $\alpha = 1$. To illustrate this type of behavior we plot the characteristic product $\Pi$ as a function of $\alpha$ (see Fig. \ref{risunokVF2}). The curve 1 corresponding to $\beta = 0$ is always above 1, grows slowly within the stability stripe, and then after reaching the triangle it starts growing more rapidly and diverges at $\alpha \to 1$. At $\beta = 0.3$ the curve 2 first goes down below 1 (instability) and only at $\alpha =0.85$ crosses the level 1.0 and enters the stability region. At large $\beta = 0.8$ the SLL phase is nearly always unstable, except for a narrow region at $\alpha \to 1$.

\section{Concluding remarks}

We have investigated stability conditions of the Sliding Luttinger liquid phase in a 2D system of parallel quantum wires with alternating coupling between nearest neighbor wires. We took  the density - density interaction into account as the principle coupling mechanism between the wires and neglected the current - current interaction. The direct single electron tunneling is assumed to be suppressed. We present an analytical derivation of the CDW and SC scaling dimensions. It is shown that in the $\alpha - \beta$ parameter space there exists a region of stability where both CDW and SC processes are RG irrelevant. That stability region exists at any value of the relative coupling strength $\alpha$ from 0 to 1. However, a weak modulation $\beta$ causes an instability at small $\alpha$. It can be an indication also that disorder of inter-wire couplings may also cause an instability of an otherwise stable SLL phase. At strong coupling for $\alpha$ close to one we found a rather broad stability region comprising about 10$\%$ of the total $\alpha - \beta$ space. Correspondingly, the values of the Luttinger parameter $K$  ensuring stability lies between  0.5 to 0.65 for $\alpha = 1$ and $\beta = 0$ (the widest span).

\end{document}